\documentclass[a4paper,12pt]{article}

\usepackage{graphicx}
\usepackage{graphicx}
\usepackage{epstopdf}
\usepackage{graphicx,epsf,amsmath,amsfonts,amssymb,amsbsy,enumerate}

\newcommand{\vev}[1]{\langle#1\rangle}
\newcommand{\emat}{\end{array} \right )}
\newcommand{\vect}{\left ( \begin{array}{c}}
\newcommand{\evect}{\end{array} \right )}

\begin{document}

\begin{center}
{\large\bf The phase structure of two color QCD and charged pion condensation phenomenon}

{{T. G. Khunjua $^1$}, {K. G. Klimenko$^2$}, {Roman N. Zhokhov$^{3}$$^\P$}}

$^1${The University of Georgia, GE-0171 Tbilisi, Georgia}

$^2${Logunov Institute for High Energy Physics,
NRC "Kurchatov Institute", 142281, Protvino, Moscow Region, Russia}

$^3${Pushkov Institute of Terrestrial Magnetism, Ionosphere and Radiowave Propagation (IZMIRAN),
108840 Troitsk, Moscow, Russia}

$^\P${$^*$zhokhovr@gmail.com}
\end{center}

\centerline{\bf Abstract}
The phase structure of dense quark matter in the two color case has been investigated with nonzero baryon $\mu_B$, isospin $\mu_I$ and chiral isospin $\mu_{I5}$ chemical potentials. 
It has been shown in the mean-field approximation that there exist three dualities,  one of them between phases with spontaneous chiral symmetry breaking and condensation of charged pions, found in the three color case. The other two dual symmetries between the phase with condensation of charged pions and the phase with diquark condensation and between chiral symmetry breaking and diquark condensation phenomena.
It has been demonstrated that due to the duality properties the phase diagram is extremely symmetric and the whole phase diagram in two color case can be obtained just 
by dualities from the phase structure of three color case. This shows that the dualities are rather usefull tool. 
It is shown that chiral imbalance generates charged pion condensation phenomenon in conditions of matter in neutron stars, i. e.  electrically neutral and $\beta$-equilibrated dense quark matter.  And that diquark condensation does not prohibit the generation of the charged PC phase by chiral imbalance at least in the two color case.\\

\section{Introduction}
It is contemplated that QCD is the theory of strong interaction and it should be used for the studies of strongly interacting matter at large temperatures and baryon densities. However, the perturbation theory cannot be used in the conditions realized in heavy ion collisions or compact star matter. 
Lattice simulations of QCD, which are nonperturbative ab-initio method and are quite successful at finite temperatures, due to a notorious sign problem cannot be applied to the case of non-zero baryon density, especially at rather large values of baryon . 
Thus, there arise the interest to QCD-like models, for example, to Nambu--Jona-Lasinio (NJL)-like models \cite{njl, buballa}.

In addition to finite temperature and non-zero baryon density, there can be important various other quantities in quark matter. Tor example, isospin imbalance, the difference between the number of $u$ and $d$ quarks, is an obvious property of neutron stars. Rather recently it was demonstrated that chiral imbalance, i. e. difference between densities of left and right-handed quarks, is a quite interesting feature of quark matter \cite{Ruggieri:2016fny,andrianov, braguta}. An this led to the whole new notion of anomalous transport phenomena, for example, the so-called chiral magnetic effect \cite{fukus}.
Chiral imbalance can be accounted for by chiral chemical potential $\mu_{5}$. There is another possibility of different chiral chemical potentials for $u$ and $d$ quarks, $\mu_{u5}\neq\mu_{d5}$, and chiral isospin chemical potential $\mu_{I5}=\mu_{u5}-\mu_{5d}$.

Recently, attention is paid to two-color QCD and QCD-like models \cite{kogut,son2,weise,ramos, andersen3,brauner1,andersen2,imai,chao,Bornyakov:2020kyz}.
Although a two color and tree color cases have substantial differences, the investigations of SU(2) QCD can provide us with valuable insight about the properties of QCD at non-zero baryon density. As an additional motivation let us recall that in the QCD with two colors there is no sign problem and ab initio lattice simulations are possible. Also one can stress that two-color QCD phase diagram is rather rich and can be quite interesting by its own.

The paper is organized in the following manner.
In Sec. 2 we introduce a two color NJL model with two quark flavors ($u$ and $d$ quarks). It
contains both quark-antiquark and diquark channels of interaction and is a low energy effective model of
the two-color QCD (see \cite{weise}). Sec. III contains the discussion on three duality properties (dual 
symmetries) of the model.
In Sec. 4 the phase structure of the models is investigated.
We discuss the story of the prediction of charged pion condensation in dense quark matter in Sec. 5.

\section{Two color NJL model}

The Lagrangian of effective tow-color four-fermion NJL model with baryon $\mu_B$, isospin $\mu_I$ and chiral isospin $\mu_{I5}$ chemical potentials is 
\begin{eqnarray}
L=\bar q \Big [i\hat\partial-m_0  +{\cal M}\gamma^0\Big]q+H\Big [(\bar qq)^2+(\bar qi\gamma^5\vec\tau q)^2+
\big (\bar qi\gamma^5\sigma_2\tau_2q^c\big )\big (\overline{q^c}i\gamma^5\sigma_2\tau_2 q\big )\Big], \label{IV.1}
\end{eqnarray}

where $q\equiv q_{i\alpha}$ is a flavor ($i=u,d$) and color ($\alpha =
1,2$) doublet and a four-component Dirac spinor as well. $\hat\partial\equiv\gamma^\rho\partial_\rho$ and charge-conjugated spinors are
$q^c=C\bar q^T$, $\overline{q^c}=q^T C$, where
$C=i\gamma^2\gamma^0$. Moreover,
one uses the notations  $\vec \tau\equiv (\tau_{1},
\tau_{2},\tau_3)$ and $\sigma_2$ for usual Pauli matrices acting in the two-dimensional flavor and color spaces, respectively.

Chemical potentials are included in the Lagrangian (\ref{IV.1}) in the following term  
$$
\bar q{\cal M}\gamma^0 q=\bar q\Big [\frac{\mu_B}{2}+\frac{\mu_I}2\tau_3
+\frac{\mu_{I5}}2\gamma^5\tau_3\Big] q.
$$
It contains baryon $\mu_B$-, isospin $\mu_I$- and chiral isospin $\mu_{I5}$ chemical potentials. 
Videlicet, this model (\ref{IV.1}) is able to describe the properties of quark matter with nonzero baryon $n_B=(n_{u}+n_{d})/2\equiv n/2$, 
isospin $n_I=(n_{u}-n_{d})/2$ and chiral isospin $n_{I5}=(n_{u5}-n_{d5})/2$ densities. These quantities are thermodynamically conjugated to the following chemical potentials $\mu_B$, $\mu_I$ and $\mu_{I5}$, respectively. 
The densities $n_B$, $n_I$ and $n_{I5}$ 
can be obtained if one differentiate the thermodynamic potential (TDP) of the model (\ref{IV.1}) with respect to the 
corresponding chemical potentials.

The Lagrangian (\ref{IV.1}) without term with chemical potentials is invariant under color $SU(2)_c$ and baryon $U(1)_B$ symmetries. Moreover, in the chiral limit ($m_0=0$), it has the same Pauli-Gursey 
flavor $SU(4)$ symmetry as two-color QCD.

Now let us note that in the chiral limit ($m_0=0$) the Lagrangian (\ref{IV.1}) with the term containing chemical potentials is no longer invariant under Pauli-Gursey $SU(4)$ symmetry. 
So, apart from the color $SU(2)_c$ group, it is invariant under
$U(1)_B\times U(1)_{I_3}\times U(1)_{AI_3}$ group, where 
\begin{eqnarray}
U(1)_B:~q\to\exp (\mathrm{i}\alpha/2) q;~
U(1)_{I_3}:~q\to\exp (\mathrm{i}\alpha\tau_3/2) q;~
U(1)_{AI_3}:~q\to\exp (\mathrm{i}
\alpha\gamma^5\tau_3/2) q.
\label{2001}
\end{eqnarray}
Note also that the quantities $n_B$, $n_I$ and $n_{I5}$ are the ground state expectation values of the densities of conserved charges corresponding
to $U(1)_B$, $U(1)_{I_3}$ and $U(1)_{AI_3}$ symmetry groups and one has 
$n_B=\vev{\bar q\gamma^0q}/2$, $n_I=\vev{\bar q\gamma^0\tau^3 q}/2$ and $n_{I5}=\vev{\bar q\gamma^0\gamma^5\tau^3 q}/2$. In addition, the Lagrangian (\ref{IV.1}) is invariant under the electromagnetic $U(1)_Q$ group,
\begin{eqnarray}
U(1)_Q:~q\to\exp (\mathrm{i}Q\alpha) q,
\label{2002}
\end{eqnarray}
where $Q$ is a matrix of electric charges. 

In order to find the TDP and study the phase structure one starts from an equivalent semi-bosonized Lagrangian $\widetilde L$ with auxiliary 
bosonic fields $\sigma (x)$, $\vec\pi  =(\pi_1 (x),\pi_2 (x),\pi_3 (x))$, $\Delta (x)$ and $\Delta^* (x)$
\begin{eqnarray}
\widetilde L=\bar q \Big [i\hat\partial-m_0+{\cal M}\gamma^0 -\sigma -i\gamma^5\vec\tau\vec\pi\Big ]q-\frac{\sigma^2+\vec\pi^2+
\Delta^*\Delta}{4H}-\frac{\Delta}{2}\Big [\bar qi\gamma^5\sigma_2\tau_2q^c\Big ]-
\frac{\Delta^*}{2}\Big [\overline{q^c}i\gamma^5\sigma_2\tau_2 q\Big ], \label{IV.2}~~~~~~~~~~2
\end{eqnarray}
It is clear that the Lagrangians (\ref{IV.1}) and (\ref{IV.2}) are equivalent, as can be seen by using the equations of motion for auxiliary bosonic fields that have the form
\begin{eqnarray}
\sigma (x)=-2H(\bar qq),&~&\Delta (x)= -2H\Big [\overline{q^c}i\gamma^5\sigma_2\tau_2 q\Big ]=-2H\Big [q^TCi\gamma^5\sigma_2\tau_2 q\Big ],\nonumber\\
~\vec\pi(x)=-2H(\bar qi\gamma^5\vec\tau q),&~&
\Delta^*(x)=-2H\Big [\bar qi\gamma^5\sigma_2\tau_2q^c\Big ]=-2H\Big [\bar qi\gamma^5\sigma_2\tau_2C\bar q^T\Big ].\label{IV.3}
\end{eqnarray}
One can easily see from (\ref{IV.3}) that $\sigma(x)$ and $\pi_a(x)$ ($a=1,2,3$) are Hermitian, i.e. real, bosonic fields, and the fields
$\Delta^*(x)$ and $\Delta(x)$ are Hermitian conjugated to each other. Indeed, $(\sigma(x))^\dagger=\sigma(x)$,
$(\pi_a(x))^\dagger=\pi_a(x)$, $(\Delta(x))^\dagger=\Delta^*(x)$ and $(\Delta^*(x))^\dagger=\Delta(x)$, where
$\dagger$ denotes the Hermitian conjugation. 
Bosonic fields $\pi_3 (x)$, $\pi^\pm (x)=(\pi_1 (x)\mp i\pi_2 (x))/\sqrt{2}$ could be identified with the pion fields. If $\vev{\sigma (x)}\ne 0$ or $\vev{\pi_0 (x)}\ne 0$, then one can see that chiral symmetry is broken down and we will call this phase chiral symmetry breaking (CSB) one. 
If $\vev{\pi_{1,2} (x)}\ne 0$, then both the isospin  $U(1)_{I_3}$ and the electromagnetic $U(1)_Q$ symmetries are broken down, this phase is called charged pion condensation phase (PC). If $\vev{\Delta (x)}\ne 0$, then the baryon symmetry gets broken down spontaneously and this phase will be called baryon superfluid phase (BSF). One can see that the ground state expectation values of all bosonic fields (\ref{IV.3}) are $SU(2)_c$ invariant, hence in this model the color symmetry can not be broken down dynamically.

Upon introduction of Nambu-Gorkov field $\Psi$
\begin{equation}
\Psi=\left({q\atop q^c}\right),~~\Psi^T=(q^T,\bar q C^{-1});~~
\quad \overline\Psi=(\bar q,\overline{q^c})=(\bar q,q^T C)=\Psi^T \left
(\begin{array}{cc}
0~~,&  C\\
C~~, &0
\end{array}\right )\equiv\Psi^T Y,
\label{IV.5}
\end{equation}
the semi-bosonised Lagrangian would take a form
\begin{eqnarray}
\widetilde L=-\frac{\sigma^2+\vec\pi^2+\Delta^*\Delta}{4H}+\frac 12\Psi^T(YZ)\Psi, \label{IV.12}
\end{eqnarray}
where matrix $Y$ is given in Eq. (\ref{IV.5}) and
\begin{equation}
Z=\left (\begin{array}{cc}
D^+, & K\\
K^*~~, &D^-
\end{array}\right )\equiv \left (\begin{array}{cc}
i\hat\partial-m_0+{\cal M}\gamma^0 -\sigma -i\gamma^5\vec\tau\vec\pi, & -i\gamma^5\sigma_2\tau_2\Delta\\
~~~~~~~~-i\gamma^5\sigma_2\tau_2\Delta^*~~~~~~~~~~~~, &i\hat\partial-m_0-\gamma^0{\cal M} -\sigma -i\gamma^5(\vec\tau)^T\vec\pi
\end{array}\right ).\label{IV.13}
\end{equation}
In the mean-field (or one fermion loop) approximation, the effective action ${\cal S}_{\rm
{eff}}(\sigma,\vec\pi,\Delta,\Delta^{*})$ can be expressed in terms of the path integral over quark fields
\begin{eqnarray}
\exp(i {\cal S}_{\rm {eff}}(\sigma,\vec\pi,\Delta,\Delta^{*}))=
  N'\int[d\bar q][dq]\exp\Bigl(i\int\widetilde L\,d^4 x\Bigr),\label{IV.14}
\end{eqnarray}
and if one express it in a more useful form
\begin{eqnarray}
&&{\cal S}_{\rm {eff}}
(\sigma,\vec\pi,\Delta,\Delta^{*})
=-\int d^4x\left [\frac{\sigma^2(x)+\vec\pi^2(x)+|\Delta(x)|^2}{4H}\right ]+
\widetilde {\cal S}_{\rm {eff}},
\label{IV.15}
\end{eqnarray}
where the quark contribution is
\begin{equation}
\exp(i\tilde {\cal S}_{\rm {eff}})=N'\int [d\bar
q][dq]\exp\Bigl(\frac{i}{2}\int\Big [\Psi^T(YZ)\Psi\Big ]d^4 x\Bigr).
\label{IV.16}
\end{equation}
One can express the path integration to $\Psi$ field
\begin{equation}
\exp(i\tilde {\cal S}_{\rm {eff}})=
  \int[d\Psi]\exp\left\{\frac i2\int\Psi^T(YZ)\Psi
  d^4x\right\}=\mbox {det}^{1/2}(YZ)=\mbox {det}^{1/2}(Z),\label{IV.17}
\end{equation}
One can define from ${\cal S}_{\rm {eff}}$ the thermodynamic potential (TDP) $\Omega(\sigma,\vec\pi, \Delta, \Delta^{*})$
in the mean-field approximation
$$
{\cal S}_{\rm {eff}}~\bigg
|_{~\sigma,\vec\pi,\Delta,\Delta^{*}=\rm {const}}
=-\Omega(\sigma,\vec\pi,\Delta,\Delta^{*})\int d^4x.
$$

The ground state expectation values of
$\vev{\sigma(x)}\equiv\sigma,~\vev{\vec\pi(x)}
\equiv\vec\pi,~\vev {\Delta(x)}\equiv\Delta,~
\vev{\Delta^{*}(x)}\equiv\Delta$, can be found as solutions of the following equations (the so-called gap equations)
\begin{eqnarray}
\frac{\partial\Omega}{\partial\pi_a}=0,~~~~~
\frac{\partial\Omega}{\partial\sigma}=0,~~~~~
\frac{\partial\Omega}{\partial\Delta}=0,~~~~~
\frac{\partial\Omega}{\partial\Delta^{*}}=0.~~~~~~~~~~~20
\label{IV.20}
\end{eqnarray}
It is assumed that $\sigma,\vec\pi,\Delta,\Delta^*$ do not depend on the space coordinates $x$.

In the mean field approximation the expression for the TDP can be obtained and it has the following form
\begin{eqnarray}
\Omega(M,\vec\pi,\Delta,\Delta^{*})
&=&\frac{(M-m_0)^2+\vec\pi^2+|\Delta|^2}{4H}+i\int\frac{d^4p}{(2\pi)^4}\ln\det L(p).
\label{IV.35}
\end{eqnarray}
The expressions for  $\det L(p)$ can be obtained analytically. 
The integration over $\vec p$ is performed up to the $|\vec p|<\Lambda$,  i. e. sharp momentum cutoff with $\Lambda=657$ MeV. The values of $H$ is taken to be $H=7.23$ GeV$^{-2}$ \cite{brauner1,andersen2}. The chiral limit, $m_0=0$, will be used to simplify numerical calculations.

\section{Duality properties}

\subsection{The case with $\mu\ne 0$, $\nu\ne 0$ and $\nu_5\ne 0$}
\label{III.B}

Above we considered the TDP as a function of all six condensates, i. e.
$M$, $\vec\pi$, $\Delta$ and $\Delta^{*}$, but if one consider the symmetries of the model the number of condensates could be reduced. As it has been discussed above, in the chiral limit and without chemical potentials, i. e.  if all of them are equal to zero, the Lagrangian is invariant with respect to $SU(4)\times U(1)_B\times SU(2)_c$ group. 
If chemical potentials are  non-zero then the Lagrangian is invariant with respect to reduced symmetry group
$U(1)_B\times U(1)_{I_3}\times U(1)_{AI_3}\times SU(2)_c$.
So in this case the TDP depends only on the following combinations $|\Delta|^2$,
$\pi_1^2+\pi_2^2$ and $M^2+\pi_0^2$,
and without loss of generality one can put $\pi_2=0$. Hence, below the TDP is assumed to depend on $M,\pi_1$ and 
Moreover, it can be shown that if $m_0\ne 0$ the global minimum point (GMP) of the TDP always have $\pi_0=0$.  
Hence, below we assume that the TDP depends only on $M,\pi_1$ and $|\Delta|$.

It can be shown with the use of any analytical calculation program that the expression for the TDP is invariant under the so-called dual transformation ${\cal D}_1$,
\begin{eqnarray}
{\cal D}_1: ~~~~\mu\longleftrightarrow\nu,~~~\pi_1\longleftrightarrow |\Delta|.
\label{IV.55}
\end{eqnarray}
This dual property of the TDP was noticed for the first time in the framework of the two-color NJL model in \cite{son2,andersen2} but for the case of $\nu_5=0$.
In particular, it was shown that the PC and BSF phases are arranged symmetrically on the $(\mu,\nu)$-phase diagram 
of the model.

Furthermore, it has been obtained that
the TDP is invariant also with respect to the following dual transformations ${\cal D}_2$ and ${\cal D}_3$
\begin{eqnarray}
{\cal D}_2: ~~\mu\longleftrightarrow\nu_5,~~M\longleftrightarrow |\Delta|;~~~~~~{\cal D}_3:
~~\nu\longleftrightarrow\nu_5,~~M\longleftrightarrow \pi_1.
\label{IV.60}
\end{eqnarray}

\section{Phase diagram}

Now let us discuss the arguments why there is no mixed phase in the phase diagram in our case and hence one can assume that the global minimum point (GMP) $(M,\pi_1,|\Delta|)$ of the TDP has only one nonzero coordinate.
There is a rather cogent argument for absence of mixed phases and it is based on dual properties discussed above. 
 The phase diagram concerning chiral symmetry breaking and charged pion condensation phenomena has the same structure in two color and three color cases in framework of NJL models.  These phenomena in the three color case has been investigated in \cite{kkz2,kkz,kkz1+1} and it has been displayed that (I) chiral and charged pion condensates do not take non-zero values simultaneously (meaning no mixed phase). Hence, this holds also for two color NJL model. Then, in order to show that (II) there is no such a region, where chiral and diquark condensates are non-zero simultaneously, the duality ${\cal D}_{2}$, which turns pion condensate to the diquark one, should be used.
 The absence of mixed phase with non-zero pion and diquark condensates can be shown applying either ${\cal D}_{2}$ to the former case (I) or the ${\cal D}_{3}$ to the latter one (II).
This does not show that there could not be a phase with all three non-zero condensates (chiral, pion and diquark), but it seems to be quite unlikely possibility.
Without dualities showing the absence of mixed phases would be rather hard and can be done only numerically.

Let us talk about possible phases in the system that can be realized. 
(i) the chiral symmetry breaking (CSB) phase: GMP has the form $(M\ne 0,\pi_1=0,|\Delta|=0)$. (ii) charged pion condensation (PC) phase: GMP of the form $(M=0,\pi_1\ne 0,|\Delta|=0)$.
(iii) baryon superfluid (BSF) phase: GMP is $(M=0,\pi_1=0,|\Delta|\ne 0)$. (iv) symmetrical phase: GMP has the form $(M=0,\pi_1=0,|\Delta|=0)$.

\subsection{The case of one chemical potential}

The phase diagram of the two color NJL model with only nonzero baryon chemical potential (although with nonzero temperature $T$) has been investigated in \cite{weise}. It was shown that at non-zero (although not for too large) $\mu$ and zero temperature $T=0$ there is the BSF phase. If one apply to this phase diagram the duality transformation ${\cal D}_1$ one would get the phase diagram with only nonzero isospin chemical potential  $\nu$. And one can see that at $\nu>0$ there is the charged PC phase. 
Now acting by the other duality transformation ${\cal D}_2$ upon the phase diagram  with nonzero $\mu$ (original one) one would get the phase structure at only non-zero $\nu_5$ and when $\nu_5>0$ observe CSB phase is realized in the system. 
One can conclude that there exist one-to-one correspondence between different phenomena, chiral symmetry breaking, charged pion condensation and diquark condensation, and various chemical potentials, chiral isospin chemical potential $\nu_5$, isospin chemical potential $\nu$ and quark chemical potential $\mu$.

\subsection{The case of two chemical potentials}
The $(\mu,\nu)$-phase diagram at $\nu_5=0$ has been already considered in \cite{son2,andersen2}. The obtained phase diagram is depicted in Fig. 1(a) and one can note that it is self-dual, i. e. BSF and charged PC phases are mirror symmetrical with respect to the $\mu=\nu$ line. This is due to the duality ${\cal D}_1$ of TDP

If one employs the dual transformation ${\cal D}_2$ and act on this diagram, one obtain the $(\nu_5,\nu)$-phase diagram of the model at $\mu=0$ (it is presented in Fig. 1(b)).
In a similar fashion, acting on the diagram of Fig. 1(a) with the dual transformation ${\cal D}_3$, one can get the $(\mu,\nu_5)$-phase diagram of the model at $\nu=0$ (see Fig. 1(c)). One can note that CSB and BSF phases are located mirror symmetrically with respect to the line $\nu_5=\mu$. Here one can also clearly see the discussed above one-to-one correspondence between different phenomena and corresponding chemical potentials.

\begin{figure}
\hspace{-1cm}\includegraphics[width=1.0\textwidth]{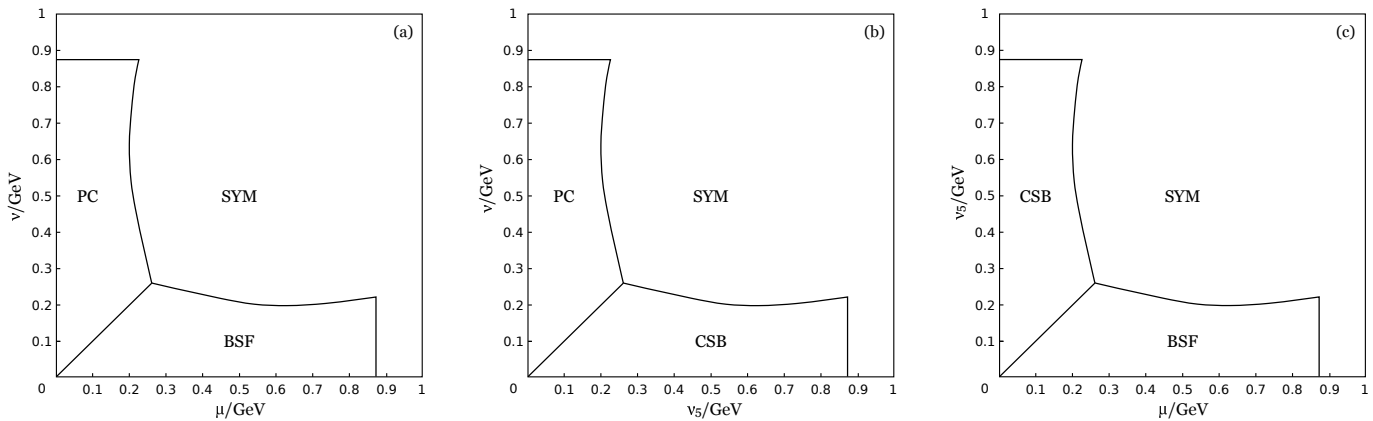}
 \caption{ (a) $(\mu,\nu)$-phase diagram at zero $\nu_{5}$. (b) $(\nu_5,\nu)$-phase diagram at zero $\mu$. (c) $(\mu,\nu_5)$-phase diagram at zero $\nu$.
}
\end{figure}

\begin{figure}
\hspace{-1cm}\includegraphics[width=1.0\textwidth]{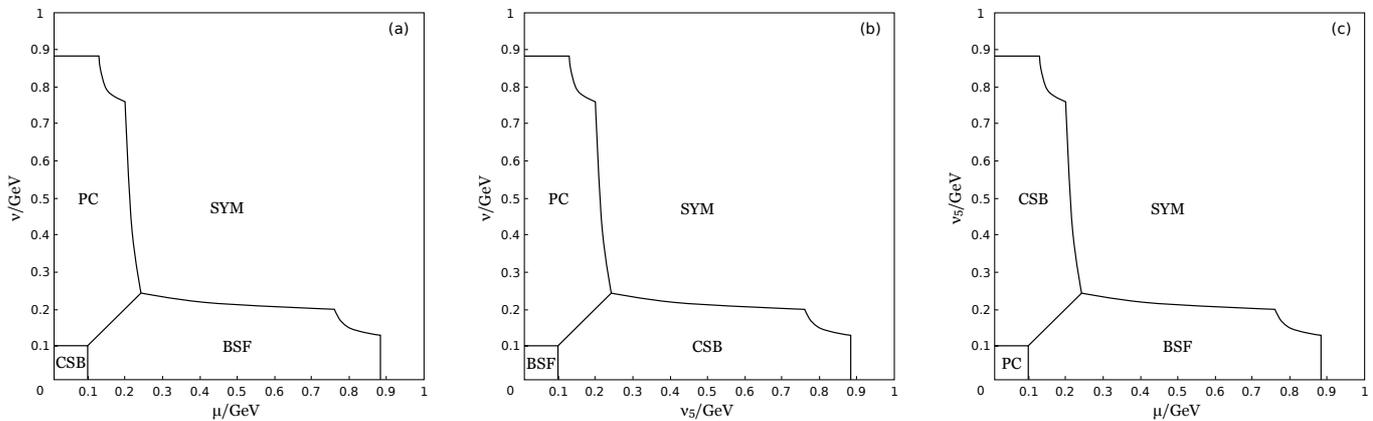}
 \caption{ (a) $(\mu,\nu)$-phase diagram at $\nu_{5}=0.1$ GeV. (b) $(\nu_5,\nu)$-phase diagram at $\mu=0.1$ GeV. (c) $(\mu,\nu_5)$-phase 
 diagram at $\nu=0.1$ GeV. All the notations are the same as in previous Fig. 1.
}
\end{figure}

\subsection{Phase structure of the general case: $\mu\ne 0, \nu\ne 0$ and $\nu_5\ne 0$}

In this section let us consider the general case when all three chemical potentials $\mu$, $\nu$, $\nu_5$ are non-zero, i. e.  $(\mu,\nu,\nu_5)$-phase diagram of the model. 
One can use the assumption that there is no mixed phase, which was proved (excluding three nonzero condensates case) above, and perform numerical calculations only for projections of the TDP on the axis of condensates ($M$, $\pi$) (one need only to find minima of the projections and compare them).

There is another way, which is even much simpler, one can use the dual properties of the model. Let us first note that chiral symmetry breaking and pion condensation phenomena in the framework of NJL model are very similar in the cases of three and two colors, what is different is diquark condensation. The $(\nu,\nu_5)$ phase diagram at different $\mu$ as well as $(\mu,\nu)$ phase diagram at different $\nu_5$, where only chiral symmetry breaking and pion condensation phenomena were accounted for, have been investigated in the case of three colors in \cite{kkz}. Therefore, one knows the behavior of PC and CSB phases in $(\nu,\nu_5)$ and $(\mu,\nu)$ phase diagrams in the case of two colors. Then, the duality transformations ${\cal D}_1$, ${\cal D}_2$ and ${\cal D}_3$ can be used to get the information about BSF phase and hence obtain the whole phase diagram. It is quite remarkable that the whole phase structure of two color NJL model can be obtained from some results of three color case. Duality constrain the phase portrait and make it so highly symmetric that this gets possible. 

Now we can discuss the phase structure itself.
Let us start with several $(\mu,\nu)$ phase diagrams at various values of $\nu_5$ that are not that large (see Figs. 1(a), 2(a) and 3(a,b)). One can note that the diquark condensation (BSF phase) is realized in the domain, where $\mu>\nu,\nu_5$, whereas if $\nu>\mu,\nu_5$ the charged PC phase is present. This behavior qualitatively can be explained in a similar way as in \cite{andersen2}, where the case $\nu_5=0$ was considered. $\nu_5$ in our case (at phase diagrams discussed) is not very large and, for simplicity, one can put it $\nu_5\approx 0$. If $\nu>\mu$, $u$ and $\bar d$ quarks form Fermi seas and the condensation of Cooper pairs $u\bar d$ is possible, and one can see that it is charged PC phase.
Whereas, if $\mu>\nu$, then $u$ and $d$ quarks form Fermi seas and in this case the formation of Cooper pairs $ud$ and their condensation can happen and this leads to BSF phase.

Now if the values of $\nu_5$ is rather large (Fig 3(c)) the BSF phase is realized mainly in the domain, where $\mu<\nu\approx\nu_5$, and the charged PC phase in the domain, where $\nu<\mu\approx\nu_5$). This is a different behavior in comparison with the one discussed above for rather small values of $\nu_5$ Figs. 1(a), 2(a) and 3(a,b). 
This difference can be qualitatively explained by the following arguments.
The Fermi surfaces of the left and right-handed $u$ and $d$ quarks has the form
\begin{eqnarray}
\mu_{uR}=\mu+\nu+\nu_5,~~\mu_{uL}=\mu+\nu-\nu_5,~~\mu_{dR}=
\mu-\nu-\nu_5,~~\mu_{dL}=\mu-\nu+\nu_5.
\label{IV.B3}
\end{eqnarray}
At $\mu<\nu\approx\nu_5$ for the quarks $d_R$ the chemical potential $\mu_{dR}$ is negative, hence the Fermi 
sea of charge conjugated $d_R^c$ quarks can be formed, so as the Fermi sea of right-handed $u_R$ quarks, which $\mu_{uR}$ is also greater than zero. The formation of particle-hole 
Cooper pair $\bar u_R d_R^c$, which has quantum numbers of $\Delta^* (x)$, is possible and its condensation leads to appearance of BSF phase.

But to get the idea of the phase structure 
at other rather large values of $\nu_5$, one can imagine that the charged PC phase (the BSF phase) of this diagram (see Fig. 3(c)) has the form of a 
boot sole pointing by its tip to the value $\mu=\nu_5=0.4$ GeV of the $\mu$-axis (to the value $\nu=\nu_5=0.4$ GeV of the $\nu$-axis).
(And in the figure, these ``boot soles'' intersect with each other, as well as with the strip of the CSB phase along the line $\mu=\nu$.) 
Then, to find (approximately) the $(\mu,\nu)$-phase diagram at another rather high value of $\nu_5$, one should, starting from the diagram of Fig. 3(c), simply shift the ``boot sole'' of the charged 
PC phase parallel to the $\mu$-axis in a position, in which its tip is directed to the point of this axis, where $\mu=\nu_5$. In a similar 
way the BSF phase should be shifted along the $\nu$-axis. 

Note, that already in Figs. 3(a,b) one can see the beginning of the process of 
forming the charged PC and BSF phases in the form of the ``boot soles''. At rather large value of $\nu_5$, say at $\nu_5=0.24$ GeV, a small 
part of the BSF phase, a tip of the ``boot sole'', penetrates into the region where $\mu<\nu$ (see in Fig. 3(b)). And at even higher values of 
$\nu_5$ in all the $(\mu,\nu)$-phase diagrams obtained in this way, the BSF phase 
(the charged PC phase) will be located mainly in the region $\mu<\nu\approx\nu_5$ (in the region $\nu<\mu\approx\nu_5$).

\begin{figure}
\hspace{-1cm}\includegraphics[width=1.0\textwidth]{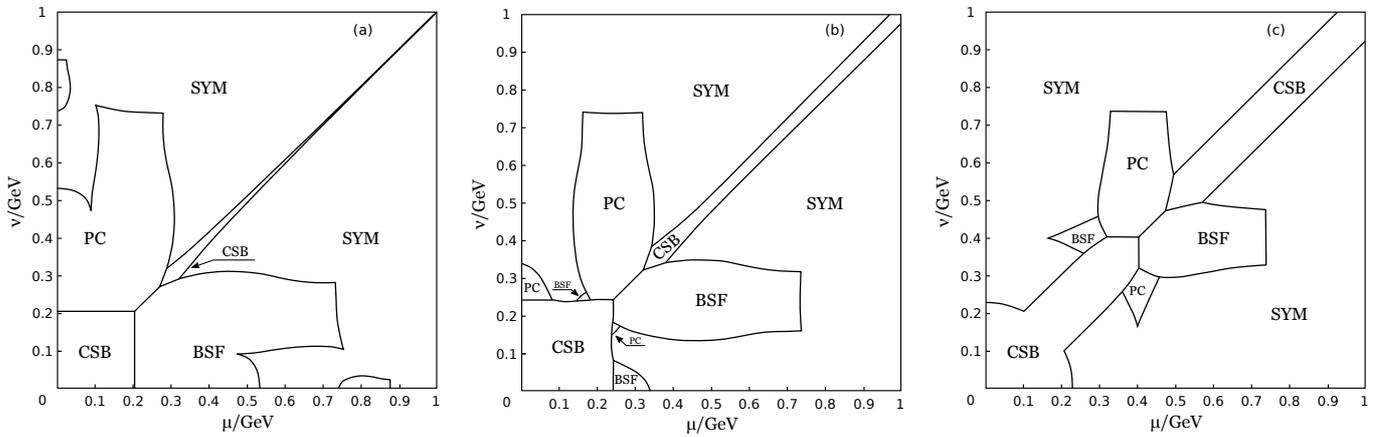}
 \caption{ (a) $(\mu,\nu)$-phase diagram at $\nu_{5}=0.2$ GeV. (b) $(\mu,\nu)$-phase diagram at $\nu_5=0.24$ GeV. (c) $(\mu,\nu)$-phase 
 diagram at $\nu_5=0.4$ GeV. All the notations are the same as in Fig. 1.
}
\end{figure}

\section{Charged pion condensation in dense quark matter}
 Let us briefly discuss the story with charged pion condensation phenomenon in {\cal dense quark matter} without chiral imbalance, i.e. at $\mu_{I5}=0$ and $\mu_5=0$. First, there has been predicted the possibility of the existence of the charged PC in dense quark matter with isospin imbalance but without the requirements of its $\beta$ equilibrium and electrical neutrality. This investigation ($\mu_B\ne 0$, $\mu_I\ne 0$, but at $\mu_Q=0$, when the contribution of electrons are not taken into account) in the chiral limit was performed in the framework of the NJL model in \cite{eklim}. It has been shown that at nonzero baryon density (and at comparatively small isospin chemical potential $\mu_I$) there is the charged PC phase in the system. Then in \cite{eklim1} it was shown that this prediction stays valid also under the conditions of neutron star matter, namely if the contribution of electrons as well as electric neutrality and $\beta$-equilibrium requirements are taken into account. The investigation was performed in terms of the NJL model and, for simplicity, the consideration, as very often is done in similar situations, has been performed in the chiral limit.

In the case of the $\beta$-equilibrated matter there is no isospin chemical potential $\mu_I$ in the Lagrangian but there is electric charge  chemical potential $\mu_Q$. It is possible to draw the $(\mu,\mu_Q)$-phase diagram of NJL model. After that, One can then draw a curve of zero electric charge density $\vev{n_Q}$ on the phase diagram.

After that, in \cite{abuki} it was shown that if one accurately include into the NJL model considerations the electric neutrality and $\beta$-equilibrium conditions, combined with the taking physical value of the current quark mass $m_0$ (though it is rather small but non-zero) it was shown that the charged PC phase is completely washed away from the phase diagram and there is no any indication of charged pion condensation in dense isospin asymmetric matter at least in the NJL model approach (see Fig. 4(a)). Similar conclusions were made also in terms of hadronic models.

Then several years after these considerations there have been found several external physically realistic parameters that create charged PC phase in dense quark/baryonic matter or promote its the generation (see the recent review \cite{Khunjua:2019nnv}). In \cite{gkkz} it was shown, in particular, that if one allows the existence of spatially inhomogeneous condensates in the system then this phase is possible. Also it was revealed in \cite{ekkz} that if one take into account the finite size of the system this phase can also be realized. Additionally, it was shown rather recently that chiral imbalance in the system can lead to the appearance of charged PC phase in dense quark (baryonic) matter as well \cite{kkz,kkz2}.

\subsection{Charged pion condensation in electric neutral matter}

It was interesting to check if the generation of the charged PC phase can also be realized in the conditions of neutron star matter (additionally assuming chiral imbalance), namely in electrically neutral and $\beta$-equilibrated dense quark matter (at $\mu_{I5}=2\nu_5\ne 0$ and $\mu_5\ne 0$ in addition), or the electric charge neutrality condition (together with $\beta$-equilibrium condition) would completely or partially destroy this effect.

It has been shown that charged pion condensation phenomenon is triggered by chiral imbalance in dense
(i.e. at $\mu_B\ne 0$ when baryon density in nonzero) with conditions of electric neutrality (with account of electrons) and $\beta$-equilibrium both in the chiral limit (at $m_0=0$) and in the physical point ($m_0\ne 0$).
The discussed generation of charged PC can be induced even by only chiral isospin chemical potential $\mu_{I5}$ (at zero
chiral chemical potential, $\mu_{5}=0$).
But if both forms of chiral imbalance are non-zero in the system ($\mu_{5}$ and $\nu_{5}$ chemical potentials), then the generation of charged PC in dense baryon matter is a quite
inevitable phenomenon. In Fig. 4(b) the phase diagram of electrically neutral and $\beta$-equilibrated dense quark matter is depicted and one can see that the neutrality line $n_{Q}=0$ intersects the huge chunk of the PC$_d$ phase.

\begin{figure}
\hspace{-1cm}\includegraphics[width=1.0\textwidth]{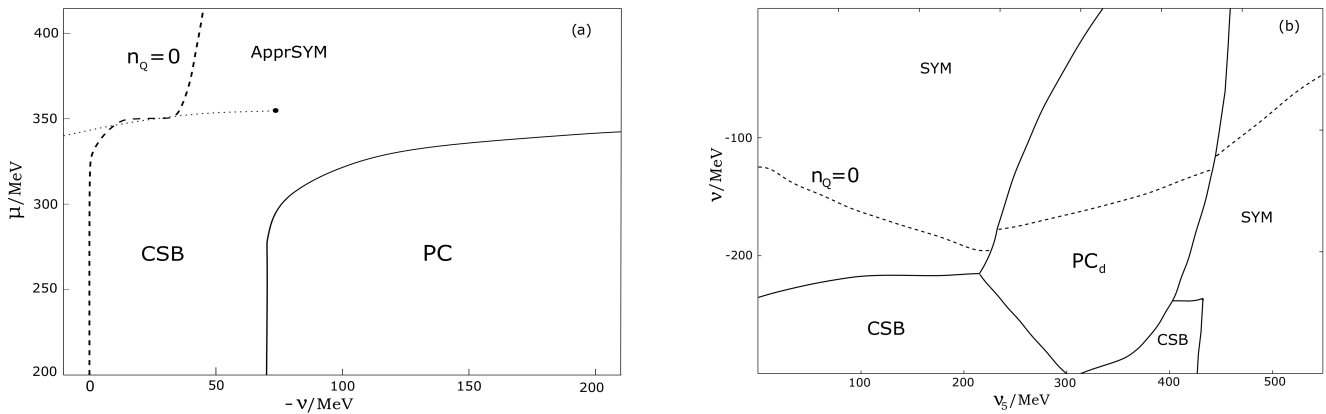}
 \caption{(a) $(\mu,\nu\equiv\mu_{Q}/2)$-phase diagram of quark matter in the $\beta$-equilibrium with electrons at $\nu_{5}=0$ and $\mu_{5}=0$. (b) $(\nu_5,\nu)$-phase diagram at $\mu=400$ MeV and $\mu_{5}=300$ MeV. 
PC denotes the phase with charged pion condensation (in PC$_d$, d means non-zero baryon density). CSB denotes the phase where there is no pion condensation and dynamical quark mass is rather large. SYM and ApprSYM are phases where quark mass is zero or around the current quark mass value, respectively.
The dashed line is "neutrality line" where the density of electric charge $\vev{n_Q}$ is equal to zero.}
\end{figure}

\subsection{Charged pion condensation in dense quark matter and diquark condensation}
It was shown chiral imbalance generate charged PC in dense baryon matter. But it is well known that in three color case in matter with rather large baryon density the color superconductivity phenomenon (diquark condensation) can take place. It starts from some value of baryon density but there is probably rather large region where both phenomena, namely generation of charged PC and diquark condensation, can exist. And there is natural question if diquark condensation phenomenon can destroy the generation of charged PC in dense baryonic matter (at least in a wide range of baryon densities).

One can note that diquark condensation appear in the two color case if there is no other chemical potentials at even infinitesimally  small values of baryon density. And probably this phenomenon is even more widespread in two color case than it is in three color case.

In the two color case we have discussed in the phase diagram of two color dense quark matter in Sec. 4 and one can see e. g. in Fig. 3(c) that diquark condensation (BSF phase) almost does not intersects with the PC$_d$ phase at all. And this means that diquark condensation does not prohibit the generation of the charged PC phase by chiral imbalance at least in the two color case.

\section{Summary}
Let us now sum up the central results of our studies of dense quark matter with isospin and chiral isospin imbalances in the framework of the two-color NJL model.

\begin{itemize}
\item It has been found that the phase structure of (two color) dense quark matter with
isospin and chiral isospin imbalances is quite rich.

It has been shown in the framework of two-color NJL model that there phase diagram of dense quark matter with
isospin and chiral isospin imbalances has a very rich structure.

\item The phenomena of chiral symmetry breaking, charged pion condensation and diquark condensation have been shown to be in a way connected with the corresponding chemical potentials, $\nu_{5}$,$\mu_I$ and $\mu_B$, respectively.

\item It has been shown that $(\mu,\nu,\nu_{5})$ phase structure of quark matter in the two color case is highly symmetric. Behind this symmetry lies the dual properties (dualities) between CSB, charged PC and BSF phenomena. One of these dualities is similar to the one found in the case of three colors.

\item The dualities are very powerful tools and can greatly simplify the studies of the phase structure. Employing only duality properties the whole phase diagram two color NJL model can be obtained from the results of previous studies of NJL model in the three color case.

\item The previously discussed generation of charged PC by chiral imbalance was shown to be a quite inevitable phenomenon and to take place also in the conditions pertinent to the neutron stars, i. e. in electrically neutral and $\beta$-equilibrated dense quark matter. 

\item It has been revealed that in dense (nonzero baryon density) quark matter the generation by chiral imbalance $\nu_{5}$ of the charged PC phase is not impeded in any way by diquark condensation at least in two color case.
\end{itemize}

\section*{Acknowledgment}
R.N.Z. is grateful for support of Russian Science 
Foundation under the grant \textnumero 19-72-00077. The work is also supported by the Foundation for the Advancement of Theoretical Physics and Mathematics BASIS grant.


\begin{thebibliography}{99}
\bibitem{njl}
Nambu Y. and Jona-Lasinio G., Phys. Rev. {\bf 122}, 345 (1961).

\bibitem{buballa}
S.~P.~Klevansky,
  Rev.\ Mod.\ Phys.\  {\bf 64}, 649 (1992);
D. Ebert, H. Reinhardt and M. K. Volkov, Prog. Part. Nucl. Phys. {\bf 33}, 1 (1994); 
T.~Inagaki, T.~Muta and S.~D.~Odintsov,
  Prog.\ Theor.\ Phys.\ Suppl.\  {\bf 127}, 93 (1997);
M. Buballa, Phys. Rep. {\bf 407}, 205 (2005); 
A. A. Garibli, R. G. Jafarov, and V. E. Rochev, 
Symmetry {\bf 11}, no. 5, 668 (2019).
  
    \bibitem{Ruggieri:2016fny}
  M.~Ruggieri, G.~X.~Peng and M.~Chernodub,
  EPJ Web Conf.\  {\bf 129}, 00037  (2016);
M.~Ruggieri and G.~X.~Peng,
  Phys.\ Rev.\ D {\bf 93},  094021  (2016).
  
   \bibitem{andrianov}
R.~Gatto and M.~Ruggieri,
  Phys.\ Rev.\ D {\bf 85}, 054013 (2012);
 L.~Yu, H.~Liu and M.~Huang,
 Phys.\ Rev.\ D {\bf 90}, 074009 (2014);
 Phys.\ Rev.\ D {\bf 94}, 014026 (2016);
 M.~Ruggieri and G.~X.~Peng,
  J.\ Phys.\ G {\bf 43}, no. 12, 125101 (2016);
  A.~A.~Andrianov, V.~A.~Andrianov and D.~Espriu,
  Particles {\bf 3}, no. 1, 15 (2020);
D.~Espriu, A.~G.~Nicola and A.~Vioque-Rodríguez,
  arXiv:2002.11696 [hep-ph].
  
  \bibitem{braguta}
V. V.~Braguta and A. Y.~Kotov,
 Phys.\ Rev.\ D {\bf 93}, 105025 (2016);
V.~V.~Braguta, E.~M.~Ilgenfritz, A.~Y.~Kotov, B.~Petersson and S.~A.~Skinderev,
Phys.\ Rev.\ D {\bf 93},  034509  (2016);
N.~Y.~Astrakhantsev, V.~V.~Braguta, A.~Y.~Kotov and A.~A.~Nikolaev,
arXiv:1902.09325 [hep-lat];
  V.~V.~Braguta, V.~A.~Goy, E.-M.~Ilgenfritz, A.~Y.~Kotov, A.~V.~Molochkov, M.~Muller-Preussker and B.~Petersson,
  JHEP {\bf 1506}, 094 (2015);
V.~V.~Braguta, M.~I.~Katsnelson, A.~Y.~Kotov and A.~M.~Trunin,
 Phys.\ Rev.\ B {\bf 100}, 085117 (2019).
 
  
  \bibitem{fukus}
K. Fukushima, D. E. Kharzeev and H. J. Warringa, Phys.Rev.D {\bf 78}, 074033 (2008).
  
    \bibitem{kogut} 
  J.~B.~Kogut, M.~A.~Stephanov and D.~Toublan,
  Phys.\ Lett.\ B {\bf 464}, 183 (1999);
  J.~B.~Kogut, M.~A.~Stephanov, D.~Toublan, J.~J.~M.~Verbaarschot and A.~Zhitnitsky,
  Nucl.\ Phys.\ B {\bf 582}, 477 (2000).

\bibitem{son2} 
  K.~Splittorff, D.~T.~Son and M.~A.~Stephanov,
  Phys.\ Rev.\ D {\bf 64}, 016003 (2001).
  

\bibitem{weise}
C.~Ratti and W.~Weise,
  Phys.\ Rev.\ D {\bf 70}, 054013 (2004).

\bibitem{ramos}
  D.~C.~Duarte, P.~G.~Allen, R.~L.~S.~Farias, P.~H.~A.~Manso, R.~O.~Ramos and N.~N.~Scoccola,
  Phys.\ Rev.\ D {\bf 93}, 025017 (2016).
  
\bibitem{andersen3} 
  J.~O.~Andersen and A.~A.~Cruz,
  Phys.\ Rev.\ D {\bf 88}, 025016 (2013).

  \bibitem{brauner1} 
  T.~Brauner, K.~Fukushima and Y.~Hidaka,
  Phys.\ Rev.\ D {\bf 80}, 074035 (2009)
  Erratum: [Phys.\ Rev.\ D {\bf 81}, 119904 (2010)].
  
\bibitem{andersen2} 
  J.~O.~Andersen and T.~Brauner,
  Phys.\ Rev.\ D {\bf 81}, 096004 (2010).
 
\bibitem{imai} 
  S.~Imai, H.~Toki and W.~Weise,
  Nucl.\ Phys.\ A {\bf 913}, 71 (2013).

\bibitem{chao} 
  J.~Chao,
  arXiv:1808.01928 [hep-ph].
  
\bibitem{Bornyakov:2020kyz} 
  V.~G.~Bornyakov, V.~V.~Braguta, A.~A.~Nikolaev and R.~N.~Rogalyov,
  arXiv:2003.00232 [hep-lat].
  
  \bibitem{kkz2}
T.~G.~Khunjua, K.~G.~Klimenko and R.~N.~Zhokhov,
Eur.\ Phys.\ J.\ C {\bf 79}, 151 (2019);
  JHEP {\bf 1906}, 006 (2019);
J.\ Phys.\ Conf.\ Ser.\  {\bf 1390}, no. 1, 012015 (2019)
  [arXiv:1812.01392 [hep-ph]].
  
  \bibitem{kkz}
 T.~G.~Khunjua, K.~G.~Klimenko and R.~N.~Zhokhov,
  Phys.\ Rev.\ D {\bf 97}, 054036 (2018).
  
  \bibitem{kkz1+1}
T.~Khunjua, K.~Klimenko and R.~Zhokhov,
EPJ Web Conf.\  {\bf 191}, 05015 (2018);
 Phys.\ Rev.\ D {\bf 98}, 054030 (2018);
  Phys.\ Rev.\ D {\bf 100}, 034009 (2019);
  Moscow Univ.\ Phys.\ Bull.\  {\bf 74}, no. 5, 473 (2019).
  
 \bibitem{eklim}
L. He, M. Jin, and P. Zhuang, Phys. Rev. D {\bf 71}, 116001 (2005);
D. Ebert and K. G. Klimenko, J.\ Phys.\ G {\bf 32}, 599 (2006);
Eur.\ Phys.\ J.\  C {\bf 46}, 771 (2006);
C.f.~Mu, L.y.~He and Y.x.~Liu,
  Phys.\ Rev.\  D {\bf 82}, 056006 (2010).
  
  \bibitem{eklim1}
D. Ebert and K. G. Klimenko,
Eur.\ Phys.\ J.\  C {\bf 46} (2006) 771.

\bibitem{abuki}
H.~Abuki, R.~Anglani, R.~Gatto, M.~Pellicoro and M.~Ruggieri,
   Phys.\ Rev.\  D {\bf 79} (2009) 034032;
 R.~Anglani,
  Acta Phys.\ Polon.\ Supp.\  {\bf 3} (2010) 735.
  
    \bibitem{Khunjua:2019nnv}
  T.~Khunjua, K.~Klimenko and R.~Zhokhov,
  Symmetry {\bf 11} (2019) no.6,  778;
   Particles {\bf 3}, no. 1, 62 (2020).
    
      \bibitem{gkkz}
  N. V.~Gubina, K. G.~Klimenko, S. G.~Kurbanov and V. C.~Zhukovsky,
  Phys.\ Rev.\ D {\bf 86}, 085011 (2012).
  
    \bibitem{ekkz}
 D.~Ebert, T. G.~Khunjua, K. G.~Klimenko and V. C.~Zhukovsky,
  Int.\ J.\ Mod.\ Phys.\ A {\bf 27}, 1250162 (2012);
  
\end{thebibliography}
\end{document}